\documentclass[%
reprint,
showpacs,preprintnumbers,
amsmath,amssymb,
aps,
prb,
bibnotes,
]{revtex4-1}
\usepackage{amsmath}
\usepackage{graphicx}
\usepackage{amssymb}
\usepackage{esint}
\usepackage{ulem}
\usepackage{cancel}

\usepackage{dsfont}

\usepackage{xcolor}
\usepackage{subfigure}
\usepackage{graphicx}
\usepackage{dcolumn}
\usepackage{bm}
\usepackage{dcolumn}
\usepackage{dsfont}
\usepackage{bbm}
\usepackage{hyperref}
\usepackage[mathlines]{lineno}
\usepackage{natbib}
\begin{document}
\title{Radiative Resistance at The Nano-scale: Thermal Barrier}
\author{N.~Zolghadr} 
\author{M.~Nikbakht}
\email{mnik@znu.ac.ir}
\affiliation{Department of Physics, University of Zanjan, Zanjan 45371-38791,Iran.}
\date{\today}
\begin{abstract}
In present article the radiative thermal current and radiative resistance are introduced and investigated in a system of parallel slabs. The system is placed in an environment with a constant temperature and subjected to a constant  temperature gradient, which causes a radiative energy flux through the system.  We have calculated the  steady-state temperatures profile of the system, assuming that the material and thickness of the middle slab could be different from the other slabs.  We propose the exact formulation for calculating the thermal current and thermal resistances in both linear and nonlinear regimes. According to our results, the middle slab  acts as a thermal {\it barrier} and depending on the width of this barrier, an extreme thermal isolation is achievable. Simulation results indicate that the thermal resistance of the barrier is an increasing function of the thickness for near-field separation distances but it is virtually insensitive to the barrier width in far field regime. The long range character of the radiative heat transfer, which occurs in system with identical slabs is also discussed.
\end{abstract}
\maketitle
\section{Introduction}\label{sec1}
In recent years, many studies have been conducted on radiative heat transfer between objects with separation distances less than the thermal wavelength \cite{PhysRevB.4.3303,Pendry_1999,articleexperiment,volokitin2007near,PhysRevB.100.235450}.
Because the radiative flux at this scale violates Stephen Boltzmann's law, heat flux management by controlling geometrical parameters and system-specific features has attracted much attention. Dependence on parameters is an interesting part of radiative heat transfer in two-body systems. These parameters can be either internal, such as size, shape, orientation, distance and material composition of objects \cite{doi:10.1063/1.4928430,doi:10.1063/1.3204481,SINGER201534,XU2019118432,NIKBAKHT2018164}, or an external parameter such as magnetic field in magneto-optical systems, thermal boundary conditions, the properties of surrounding media, or an electric field in metallic material systems\cite{10.1117/1.JPE.9.032711,PhysRevB.95.235428,PhysRevLett.118.173902,PhysRevLett.116.084301,Shuttling,2019JETPL.109..749V,Volokitin_2020,VolokitinPersson}. The quantitative form of the radiative heat flux can change as parameters are varied. In particular, the heat flux can be enhanced or decreased, or the net direction can change. As we move up from two-body to three-body systems, the radiative heat transfer can be tuned by changing the parameters, but theoretically and more recently it has been shown experimentally that  three-body systems can provide the possibility to enhance radiative heat transfer over two-body counterparts \cite{ben2011many,doi:10.1063/1.4894622,PhysRevB.97.075422,thompson2019nanoscale,PhysRevB.99.035433,PhysRevA.89.052104,PhysRevB.95.125411,PhysRevB.95.125411,PhysRevB.97.165437}. There has been a large
amount of literature seeking to improve thermal rectification, thermal switching and thermal splitting by controlling various material and structural parameters\cite{GU2015429,doi:10.1063/1.4964698,SONG201980}. As the number of objects in the system increases, the many-body effects become very significant \cite{PhysRevB.96.125436,PhysRevMaterials.3.015201,DONG2017114,PhysRevB.99.201406,PhysRevB.100.205422,SONG2020119346} and as expected it influences the dynamics of temperature \cite{Nikbakht_2015,dynamicstem}. On the other hand, the dynamics  and the steady-state radiative heat flux may exhibit sensitive dependence on parameters, initial conditions and also on the thermal boundary conditions \cite{dynamicalresponce,PhysRevB.90.045414,PhysRevB.101.041409}. Recent  theoretical  work on the thermal bistability has  highlighted  the importance of an external heat flux on the thermal switching  in near-field  radiative   heat   transfer \cite{PhysRevB.101.041409}.
It has been previously established  that  the   many-body parallel planar systems can provides  distinctive properties for significant enhancement of near-field heat transport \cite{PhysRevB.97.035423,articlelayyered}, and the geometrical properties as well as the initial condition for temperatures can have a remarkable effect on the temperature evolution \cite{4bodydynamics}.  

The transfer of large amounts of energy between system components in the near-field regime results in a very strong temperature coupling at these scales. However, finding ways to minimize radiation heat transfer in systems that require thermal insulation is particularly important. In {\it conductive} heat transfer this isolation is mainly carried out using multi-layer structures \cite{TIEN1973349,koebel:hal-00700709,PhysRevB.95.155309}. The temperature profile in such structures does not show a monotonic trend across an interface between different materials. Instead, there is a temperature difference athwart the boundaries. Similarly, a system of planar objects can resist radiation heat flux and it is expected to cause discontinuities in temperature profile. In other words, the components of the system provide a thermal resistance that must be considered in thermal design or analysis.

In this article, we have investigated the radiative thermal current and steady state temperature profile in a parallel planar system which is subjected to an external temperature gradient and transfer heat in the form of radiation.  Using a simple example it is shown that the steady-state {\it radiative thermal thermal current} can be tuned by engineering the intrinsic properties of each
layer. Moreover, inspired by the idea of Kapitza resistance and interfacial thermal resistance \cite{RevModPhys.41.48,kapitza,MAITI1997517}, we have introduced a radiative thermal resistance in parallel planar objects and demonstrate the possibility of extreme radiative thermal isolation. We have introduced {\it radiative thermal barrier} and the linear and non-linear resistances are calculated for barriers with different materials.  It is shown that both thermal current and temperature profile in steady-state regime depend strongly  on  the  width and the composition of the barrier. The numerical results indicate that for a given composite system, due to the existence of a thermal barrier, the radiative thermal resistance depends not only on the width of the barrier but also on the separation distances. While we have used Silicon Carbide (SiC) and hexagonal Boron Nitride (hBN) as typical materials for slabs, the proposed formalism is general and can be applied to any planar system with arbitrary parameters (materials, widths, vacuum gap distances).  Moreover, the proposed quantities (thermal resistance and radiative current) can be used to analyze the results of many studies in the field of radiative heat transfer and improve our understanding of the subjects in this field. 

The structure of the paper is as follow. The formalism is developed in Sec.~\ref{sec2}.  In Sec.~\ref{sec3}  we start with a simple prototypical example of a thermal barrier. We have computed the temperature profile in an array of Silicon Carbide (SiC) slabs when a hexagonal Boron Nitride (hBN)  or SiC slab is embedded in the middle of the system Sec.~\ref{sec3a}. The radiative thermal resistances are calculated in Sec.~\ref{sec3b}. The effect of the barrier thickness on the temperature profile and thermal resistances are investigated in Sec.~\ref{sec3c}. In Sec.~\ref{sec3d} the thermal resistance of the barrier compared for  near-field and far-field regime. Finally, this study is summarized in Sec.~\ref{sec4}.

\begin{figure}
\includegraphics[height=6cm,width=8cm]{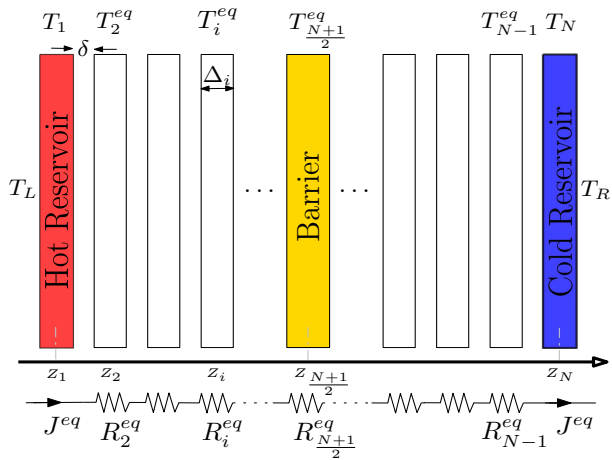}
\caption{(Color online) Schematic drawing of a thermal barrier of parallel slabs with width $\Delta$ separated by vacuum gap $\delta$. The two external slabs (reservoirs) are kept at fixed temperatures $ T_{1}=400$~K and $T_{15}=300$~K  Circuit diagram of the parallel planar system; $R_i^{eq}$ denotes a resistance, and $J^{eq}$ represents the radiative thermal current in the steady-state regime.}
\label{Figure.1}
\end{figure}
\section{PHYSICAL SYSTEM AND MODEL}\label{sec2}
The schematic of the system under consideration is shown in Fig.~\ref{Figure.1}. It consists of N parallel slabs with thicknesses $\Delta_i$ separated by vacuum gaps of width $\delta$. The slabs are along the z-axis at positions $z_i$. The system is in thermal baths ({\it environment}) from left and right at fixed temperatures $T_L\equiv T_0$ and $T_R\equiv T_{N+1}$, respectively.  Moreover, the first ($i=1$) and the last ($i=N$) slabs  are connected to {\it reservoirs} with fixed temperatures $T_1$ and $T_N$, respectively. Based on the boundary condition of the system, i.e. $(T_L,T_1,T_N,T_R)$, the radiative heat transfer that takes place along the system, drives the system from an initial non-equilibrium state $[T_L,T_1,T_2(0),\cdots,T_{N-1}(0),T_N,T_R]$ to a  non-equilibrium steady states configuration $[T_L,T_1,T_2^{eq},\cdots,T_{N-1}^{eq},T_N,T_R]$, in which each slab is in local thermal equilibrium and  there is no time variation of physical variables. The temporal behaviour of each  slab is governed by equation:
\begin{equation}
 \rho_ic_i\Delta_i\dfrac{\partial T_i}{\partial t}=\varphi_i(T_L,T_1,\cdots,T_N,T_R)~~i=2,\cdots,N-1
\label{eq1}
\end{equation}
where $\Delta_i$ is the thickness of the slab with mass density $\rho_i$ and heat capacity $c_i$. For a given thermal boundary condition and initial condition for temperatures, the net radiative heat flux per unit surface received by the i-th slab is given by \cite{4bodydynamics}:
\begin{equation}
\label{eq2}
\varphi_{i}=\sum_{j=0}^{N+1}\left[\int_{0}^{{\infty}}\frac{d\omega}{2\pi}\int_{0}^{{\infty}} \dfrac{dk}{2\pi}k\sum_{p=\{S,P\}}\Theta(\omega,T_j)\mathcal{T}^{j,i}(\omega,k,p)\right],
 \end{equation}
for $i=2,\cdots,N-1$. In this relation, the second summation runs over two physical polarization states of the radiating field, i.e. S and P polarizations respectively.  Moreover, $ \Theta(\omega,T_j)=\hbar\omega/(e^{\hbar\omega/k_{B}T_j}-1) $ denotes the mean energy of Plank oscillator at temperature $T_j$. The contribution of each slab  in the heat flux is given by the transmission coefficient $ \mathcal{T}^{j,i}=\mathcal{T}^{j,i}(\omega,k,p)$, which depends on the geometrical and intrinsic features of the system (see the appendix for more detail). 
The solution of Eq.~(\ref{eq1}) can be visualized as a trajectory in a (N-2)-Dimensional phase space. However, we are only interested in the long-time behavior of the system, i.e. the steady-state temperature profile that the system is able to reach as $t\rightarrow\infty$. Since the right-hand-side of Eq.~(\ref{eq1}) does not depend on $t$ explicitly, the system is autonomous and we only need to find fixed point of the system \cite{strogatz:2000}. The steady state temperature profile of Eq.~(\ref{eq1}) is defined by the fact that the net energy flux on each slab vanishes, i.e. $\varphi_i(\textbf{T}^*)=0$ for $i=2,\cdots,N-1$, where $\textbf{T}^*\equiv (T_L,T_1,T^{eq}_2,\cdots,T_{N-1}^{eq},T_N,T_R)$ is the fixed point of the system in phase space. For further investigation of the steady-state of the  system, we summarized and extend Eq.~(\ref{eq2}) to cover all system components, including slabs, environments and reservoirs. Hence
\begin{equation}
\varphi_i=F_{i,i}+\sum_{j\neq i}F_{j,i}+F^{ext}_i,
\label{eq3}
\end{equation}
for $i=0, 1, \cdots, N, N+1$. In second term the value of index $j$ runs from $0(L)$ to $N+1(R)$ (including the reservoirs and the external environment). Here $F_{i,i}\leqslant 0$ is a radiative cooling of the i-th component which could be a slab, a reservoir or an environment. Moreover, $F_{j,i}\geqslant0$ is the power transferred from the j-th component to the i-th one. Finally, an external amount of heat ($F^{ext}$) should be transfer from, or to, heat reservoirs and environments in order to keep them in constant temperatures. It should be emphasized that $F_i^{ext}=0$ for slabs with varying temperature, i.e. $i=2,\cdots,N-1$  to match Eq~(\ref{eq2}). 

Suppose the left environment absorbs the rate of heat $F_L^{ext}$ and first slab absorbs the rate of heat $F_1^{ext}$. According to the  conservation of energy $\sum_{i=0}^{N+1}\varphi_i=0$, it is easy to show that $F_L^{ext}(t\rightarrow\infty)+F_1^{ext}(t\rightarrow\infty)=-[F_R^{ext}(t\rightarrow\infty)+F_N^{ext}(t\rightarrow\infty)]$, which can be referred to as the left-right symmetry at steady state. Specifically, we can say now that the same amount of heat which is given to the left environment and reservoir at a time  is equal to the one that taken from the right environment and reservoir, or visa versa. By setting $\varphi_i=0$ in Eq~(\ref{eq3}) for $i=0,1$ and using Eq.~(\ref{eq2}), it is straightforward to show that 
\begin{eqnarray}
\label{eq4}F_1^{ext}&=&-\sum_{j=0}^{N+1}\int_{0}^{{\infty}}\frac{d\omega}{2\pi}\int_{0}^{{\infty}} \dfrac{dk}{2\pi}k\sum_{p}\Theta(\omega,T_j)\mathcal{T}^{j,1},\\
\label{eq5}F_L^{ext}&=&-\sum_{j=0}^{N}\int_{0}^{{\infty}}\frac{d\omega}{2\pi}\int_{0}^{{\infty}} \dfrac{dk}{2\pi}k\sum_{p}\Theta_{j,R}\mathcal{T}^{L,j},
\end{eqnarray}
with $\Theta_{j,R}=\Theta(\omega,T_j)-\Theta(\omega,T_R)$. It is clear that these external powers are time-dependent during the initial stage of the dynamics of temperatures in the system. However, they eventually approach the steady-state values as the system reaches local thermal equilibrium. In addition, we know that due to the temperature difference caused by  the boundary conditions,  the {\it radiative thermal current} flows along the z direction either to the left or to the right. Hence, we define the net current flow along the system as: 
\begin{equation}
J^{eq}=F^{ext}_L+F^{ext}_1\equiv -(F^{ext}_R+F^{ext}_N),
\label{eq6}
\end{equation}
which remains constant in the steady-state regime. Here, $J^{eq}$ represents the radiant energy passing through the system in local thermal equilibrium. It is important to emphasize that for the system under consideration, the transmission probabilities do not depend on temperature as in phase changed materials, which implies that the power dissipated in each slab ($\varphi_i$) is a continues function of temperatures and the off-diagonal elements in the Jacobian matrix of the system ($\frac{\partial \varphi_i}{\partial T_j}, i\neq j$) have constant sign, independent of the system's state. 

As a result, the system of equations $\varphi_i=0$ have only one fixed point and since the system is non-conservative, the fixed point is stable. On the other hand, the steady-state and so the equilibrium thermal current do not depend on the choice of initial condition of the system, i.e. [$T_2(0),\cdots,T_{N-1}(0)$]. Using Eqs.~(\ref{eq4}) to (\ref{eq6}), the steady-state thermal current passing through the system can be expressed as
\begin{equation}
J^{eq}=\int_{0}^{{\infty}}\frac{d\omega}{2\pi}\mathcal{C}^{eq}(\omega),
\label{eq7}
\end{equation}
where the monochromatic thermal current coefficient is defined as
\begin{equation}
\mathcal{C}^{eq}(\omega)=\int_{0}^{{\infty}} \dfrac{dk}{2\pi}k\mathcal{S}^{eq}(\omega,k),
\label{eq8}
\end{equation}
where $\mathcal{S}^{eq}(\omega,k)=\mathcal{S}^{eq}_P(\omega,k)+\mathcal{S}^{eq}_S(\omega,k)$ with 
\begin{equation}
\mathcal{S}^{eq}_p(\omega,k)=\sum_{j=0}^N\Theta_{R,j}^{eq}\left[\mathcal{T}^{L,j}(\omega,k,p)+\mathcal{T}^{j,1}(\omega,k,p)\right].
\label{eq9}
\end{equation}
Here, $\Theta_{R,j}^{eq}=\Theta(\omega,T_R)-\Theta(\omega,T_j^{eq})$ and $\mathcal{S}^{eq}(\omega,k)$ can be interpreted as the steady-state dispersion relation of the thermal current passing through the system. Moreover, $\mathcal{S}^{eq}_S(\omega,k)$ and $\mathcal{S}^{eq}_P(\omega,k)$ are the dispersion relations of the thermal current for the s- and p-polarized modes. For the special case of $T_L=T_R$, Eq.~(\ref{eq9}) reduces to
\begin{equation}
\mathcal{S}^{eq}_p(\omega,k)=-\sum_{j=0}^{N+1} \Theta(\omega,T^{eq}_j)\left[\mathcal{T}^{L,j}(\omega,k,p)+\mathcal{T}^{j,1}(\omega,k,p)\right].
\label{eq10}
\end{equation}

It should be emphasize that, $\mathcal{S}$ depends not only on the choice of materials and system geometrical properties, but also on the steady-state temperature profile. For very large separation distances, where the far-field interaction dominates, the thermal current and so the resistances could mainly decided by the coupling of the middle slabs with the thermal baths (depending on the boundary conditions and materials) rather than the reservoirs temperature gradient. For the sake of simplicity we take $T_L=T_R=300$~K in our calculations. It is also easy to show that in the absence of thermal baths ($T_L=T_R=0$~K) we have $\mathcal{S}^{eq}_p(\omega,k)=-\sum_{j=1}^N \Theta(\omega,T^{eq}_j)\mathcal{T}^{j,1}(\omega,k,p)$, which for the special case of $N=2$,  Eq.~(\ref{eq7}) reduces to the well known formula for the heat flux exchanged between two parallel slabs \cite{doi:10.1063/1.3204481}.
 
There is an electrical analogy with radiative heat transfer that can be used to exploited in, see Fig.~\ref{Figure.1}. From this perspective the radiative heat flux is equivalent to the electric current and each slab is a pure resistance to radiative heat flux. Hence, we can define slab resistance as
\begin{equation}
 R_i=\dfrac{\Delta T_i}{J_i}, 
\label{eq11}
\end{equation}
where by definition, $\Delta T_i=(T_{i-1}-T_{i+1})/2$ is the temperature difference at the position of i-th slab and $J_i$ is the net radiative thermal current passing through it.  It is plausible that the thermal current and so the temperatures are function of times. However, when the system reaches steady-state, they depend only on the temperature of the two reservoirs and environments, the slab properties and their separation distances. Since the thermal current has only one path to take in the system under consideration, it is the same through all slab at the steady-state regime, i.e. $J_i\rightarrow J^{eq}$. It also seems reasonable to postulate that thermal resistances are independent of environment and reservoirs temperatures in linear regime where $T_L\sim T_1 \sim T_N\sim T_R$. To this end,  a more useful quantity to work with is the  {\it linear resistance} of slabs 
\begin{equation}
 R_i^{eq}=\lim_{J\rightarrow 0}\dfrac{d(\Delta T_i)}{dJ} {\Big |}_{eq}. 
\label{eq12}
\end{equation}

In the system under consideration, we have used $T_L=T_R=300$~K, since the contribution of  environments in the thermal current is small compared to that of reservoirs (i.e. $|F^{ext}_L|\ll|F^{ext}_1|$ especially for small separation distances), the derivative evaluated at thermal equilibrium in the limits of $T_1\sim T_N$. The total thermal resistance of the system could be calculated by simply adding up the resistance values of the individual resistors, i.e. 
\begin{equation}
R_{total}^{eq}=\sum_{i=1}^{N}R_i^{eq}\equiv\lim_{\Delta T\rightarrow 0}\frac{\Delta T}{J^{eq}}, 
\label{eq13}
\end{equation}
with $\Delta T=T_1-T_N$. We could also define a thermal boundary resistance, similar to the Kapitza resistance\cite{kapitza}, as the ratio of the temperature variation at the interface to the heat current across it. As a result, like a series circuit, the resistance of each slab would be the addition of its interface resistances
 \begin{equation}
 R_i^{eq}=R_{il}^{eq}+R_{ir}^{eq}, 
\label{eq14}
\end{equation}
where $R_{il}^{eq}=(T_{i-1}^{eq}-T_i^{eq})/2J_i^{eq}$ is the resistance of the left interface of the slab and $R_{ir}^{eq}=(T_i^{eq}-T_{i+1}^{eq})/2J_i^{eq}$ is the right part. It should be emphasizes that the  interface resistances $R_{il}^{eq}$ and $R_{ir}^{eq}$ are equivalent of the Kpitza resistance between interfaces, and as long as the thickness of a layer is not zero (or it is sandwiched between different media), the layer persist against the incident thermal current and has a resistance $R_i^{eq}\neq 0$.  

\section{RESULTS AND DISCUSSION}\label{sec3}

Based on the framework built above, we can calculate
the steady-state temperature profile of a typical radiative thermal barrier. The system that we consider here consists of 15 slabs, which is a hBN or a SiC slab with thickness $\Delta_8=\Delta_{barrier}$ sandwiched  between 14 identical SiC slabs.  The separation between slabs are equal, i.e. $\delta$, and the system is positioned in an environment with constant temperature $T_L=T_R=300$~K. The thermal gradient is applied by maintaining the two reservoir slabs at constant but different temperatures. For the non-linear regime, the temperature of hot reservoir and cold reservoir are fixed at $T_1 = 400$~K and $T_{15} = 300$~K, respectively. However, we have used $T_1\sim T_{15}=300$~K to calculate resistances in linear regime. For the complex dielectric function of SiC and hBN, we used the Lorentz-Drude model:
\begin{equation}
 \epsilon(\omega)=\epsilon_{\infty}\dfrac{\omega_{L}^{2}-\omega^{2}-i\Gamma\omega}{\omega_{T}^{2}-\omega^{2}-i\Gamma\omega}, 
\label{eq15}
\end{equation}
where the parameters for silicon carbide (SiC) are as follow: $\epsilon_{\infty}=6.7$ is the high frequency dielectric constant, $\omega_{L}=1.83\times 10^{14}$~rad/s is the longitudinal optical frequency, $\omega_{T}=1.49\times 10^{14}$~rad/s is the transverse optical frequency,  and $\Gamma = 1.0\times 10^{12}$~rad/s is damping coefficient. While for hexagonal Boron Nitride (hBN) these constants are,  $\epsilon_{\infty}=4.9$, $\omega_{L}=3.03\times 10^{14}$~rad/s, $\omega_{T}=2.57\times 10^{14}$~rad/s, and $  \Gamma = 1.0\times 10^{12} $~rad/s.
 
\subsection{TEMPERATURE PROFILE}\label{sec3a}
\begin{figure}
\includegraphics[height=6cm,width=8cm]{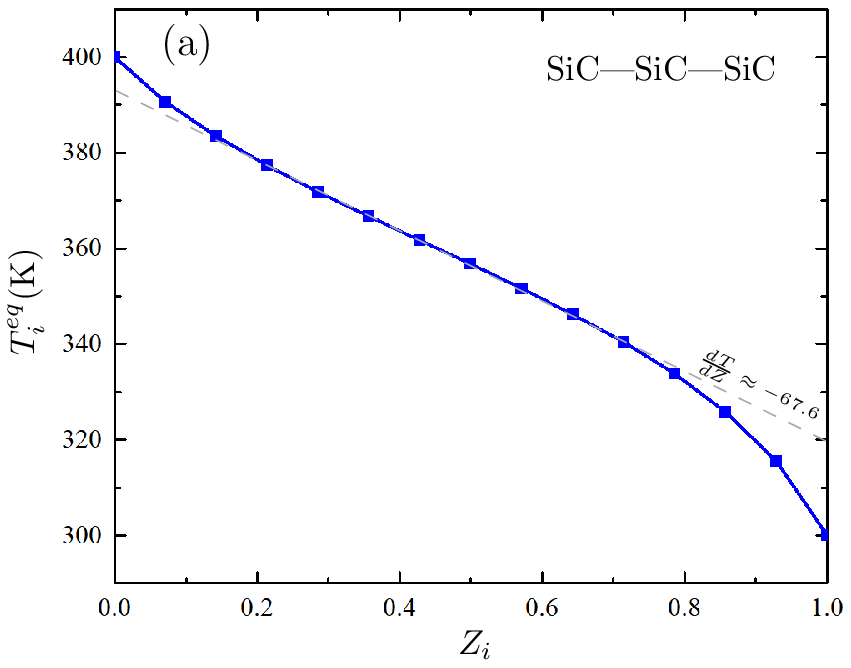}
\includegraphics[height=6cm,width=8cm]{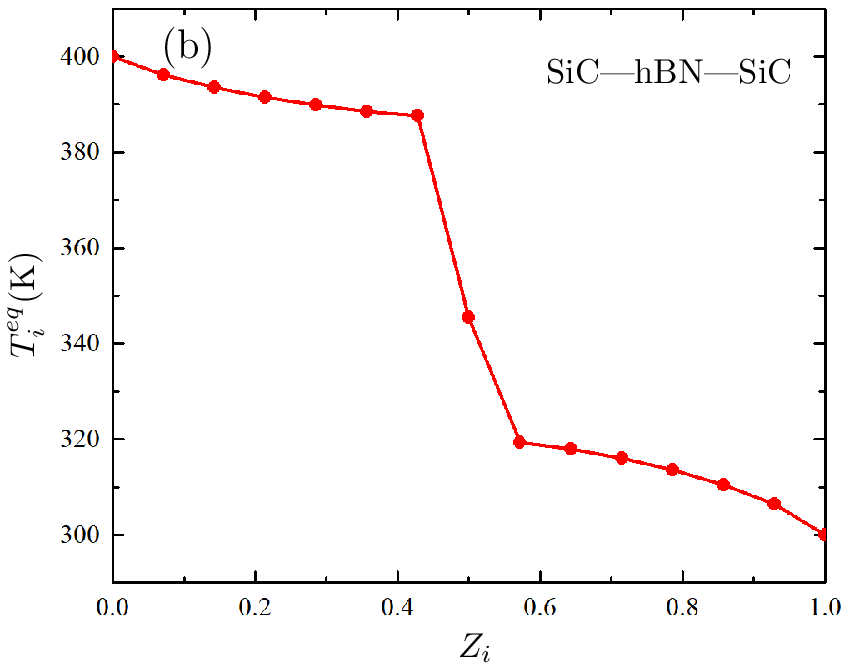}
\caption{(Color online) The steady state temperature profile for a 15-body parallel slab system, as a function of normalized position of each slab. The boundary condition is ($T_L,T_1,T_{15},T_R$)=($300,400,300,300$)~K. The thickness of slabs are the same and considered  $\Delta=200$nm and the width of vacuum gaps between slabs are $\delta=100$nm. The material used for the slabs is SiC and for the barrier is: (a) SiC, (b) hBN.}
\label{Figure.2}
\end{figure}
Figure.~\ref{Figure.2} shows the stationary state temperature profile of a  15-body parallel planar system as a function of normalized position of slabs $Z_i$. The temperature of the reservoirs and environments are ($T_L,T_1,T_{15},T_R$)=($300,400,300,300$)~K.  The slab separation distances are $\delta=100$nm and the thickness of the barrier is chosen as $\Delta_8=\Delta_{barrier}=200$nm equal to the rest of the slabs. 

The typical temperature  profile of Fig.~\ref{Figure.2}(a) which corresponds to SiC barrier (SiC--SiC--SiC system)  is clearly continuous. This profile has a part that varies roughly linearly across a large portion of the system with large gradients at the two ends due to boundary effect. Since the slabs are very close to each other, near field radiative heat transfer is the dominant mechanism that determines the steady-state temperatures. As a result, the coupling with the left and right reservoirs is strong and the temperature gradient in the profile is large.  
Figure.~\ref{Figure.2}(b) displays a similar thermal profile when the material used for the barrier (slab no. 8) is hBN (i.e. SiC--hBN--SiC system). This profile is dramatically different from the system with SiC barrier (Fig.~\ref{Figure.2}(a)) in that a sharp discontinuity in temperature appears across the barrier position. In addition, there are pronounced boundary effects as for the SiC case. It can be seen that the left slabs ($Z_i<0.5$) coupled to the left reservoir ($T_1=400$~K) and the right slabs ($Z_i>0.5$) are isolated from the left reservoir and well coupled to the right reservoir ($T_{15}=300$~K). 

\subsection{THERMAL RESISTANCE}\label{sec3b}
\begin{figure}
\subfigure[]{\includegraphics[height=6cm,width=7.7cm]{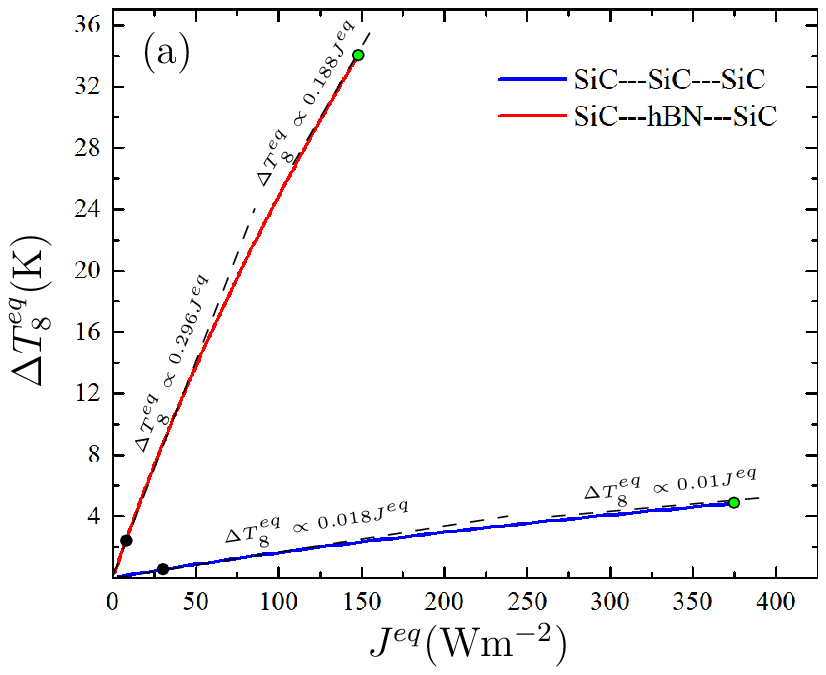}}
\subfigure[]{\includegraphics[height=6cm,width=8cm]{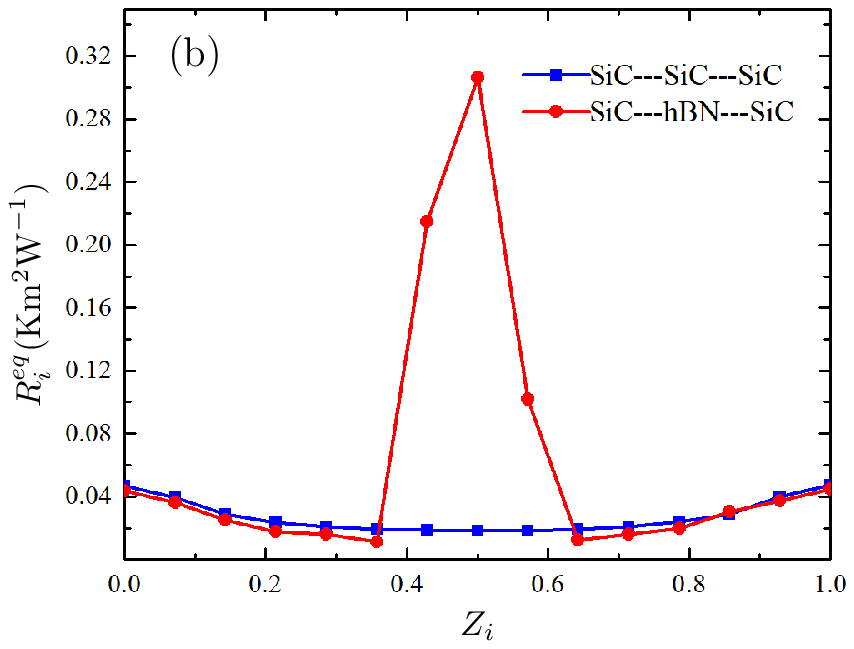}}
\caption{(Color online)  (a) Temperature drop at the barrier vs. thermal current for two considering configurations in figure~\ref{Figure.2}. Typical boundary condition that the system behaves linearly is presented by a solid black dot, i.e. ($T_L,T_1,T_{15},T_R$)=($300,310,300,300$)~K, and that the system behaves non-linearly with solid green dot ($T_L,T_1,T_{15},T_R$)=($300,400,300,300$)~K. (b) The profile of the linear resistance for a 15-body parallel slab system, as a function of normalized position of each slab.
 }
\label{Figure.3}
\end{figure}

In Fig.~\ref{Figure.3}(a) the temperature drop at the barrier $\Delta T_8^{eq}$ is plotted  against steady state thermal current $J^{eq}$ that corresponds to different boundary conditions. As it can be seen, the temperature drop increased linearly with thermal current for both SiC and hBN barriers, but for very large thermal currents, their behaviour becomes non-linear, to some extent. The linear regression is used to determine the slope that is the resistance of the barrier in both linear and non-linear regimes. As an example, the solid black dots on the diagram correspond to a linear regimes caused by  boundary condition ($T_L,T_1,T_{15},T_R$)=($300,310,300,300$)~K. On the other side, the solid green dots correspond to a non-linear regimes for boundary condition ($T_L,T_1,T_{15},T_R$)=($300,400,300,300$)~K. The temperature profiles of the latter case are those that shown in Fig.~\ref{Figure.2}. Compared  with  the  linear thermal resistance of $0.018~{\textrm  K}{\textrm m}^2{\textrm W}^{-1}$ for SiC barrier, the thermal resistance is 16 times higher for hBN barrier. It is clear from Fig.~(\ref{Figure.3}a) that larger temperature gradients $\Delta T$, create a larger temperature difference across the barrier $\Delta T_8^{eq}$ and results in higher thermal current $J^{eq}$. We observe that thermal resistances decrease with temperature in a power law form, and the decrease is larger for interface with weaker coupling (here the hBN barrier). 
\begin{figure}
\includegraphics[height=6cm,width=8cm]{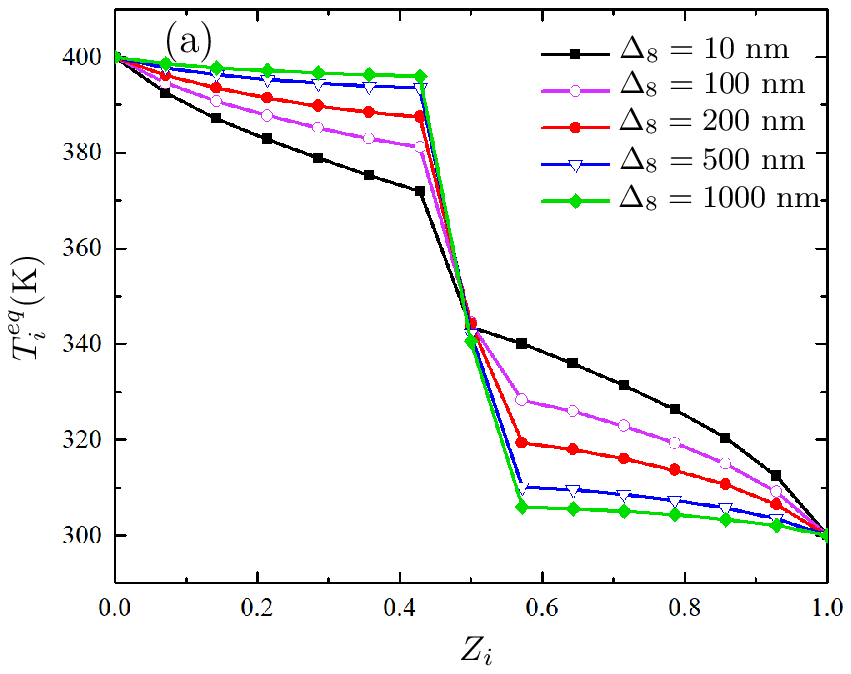}
\includegraphics[height=6cm,width=8cm]{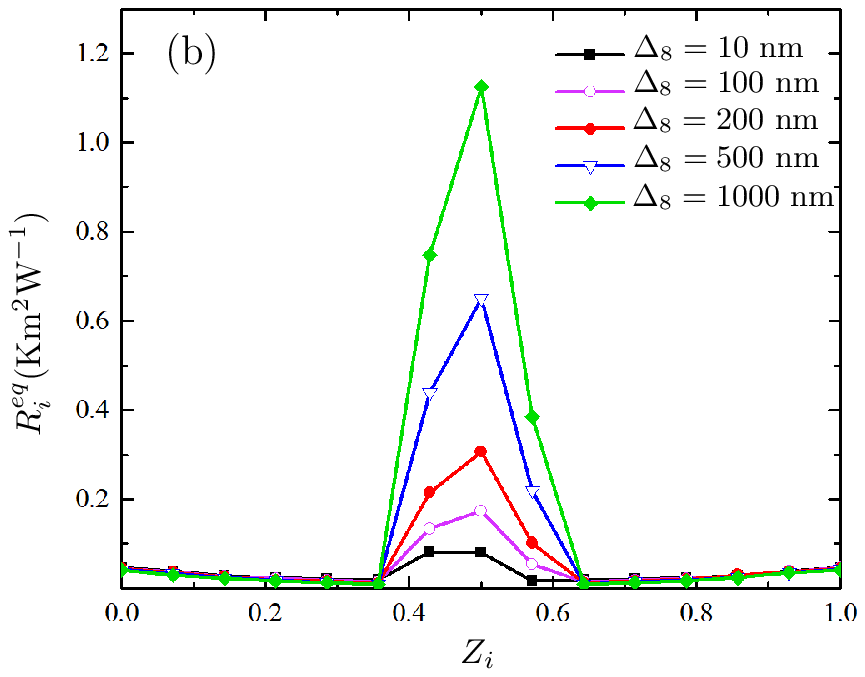}
\caption{(Color online) (a) The steady state temperature profile and (b) the corresponding linear resistance  profile for the SiC--hBN--SiC system as a function of normalized position of each slab for different hBN thicknesses. The thickness of other slabs are the same and considered $\Delta=200$nm and the width of vacuum gaps between slabs are $\delta=100$nm. }
\label{Figure.4}
\end{figure}
Using Eq.~(\ref{eq12}), we have calculated linear thermal resistance $R_i^{eq}$ of all slabs and results are shown for both SiC--SiC--SiC and SiC--hBN--SiC systems in Fig.~\ref{Figure.3}(b) as a function of normalized position of slabs. We observe that both systems show similar trends on the sides, however, the resistance of the hBN slab and its neighbours are ($\sim$ 20 times) greater in SiC--hBN--SiC system compared to the SiC barrier in SiC--SiC--SiC system. 

\subsection{BARRIER WIDTH EFFECT}\label{sec3c}
To analyze the effect of barrier width on the radiative thermal transport properties, in Fig.~\ref{Figure.4}(a) we present the temperature profiles of the SiC--hBN--SiC system for different hBN thicknesses. As can be seen, the increase in the thickness of the hBN barrier is associated with the increase in temperature discontinuity on both sides of the barrier. The corresponding linear resistance profiles which are shown in Fig.~\ref{Figure.4}(b) confirm that as the barrier thickness increases, its thermal resistance increases.
\begin{figure*}
\includegraphics[]{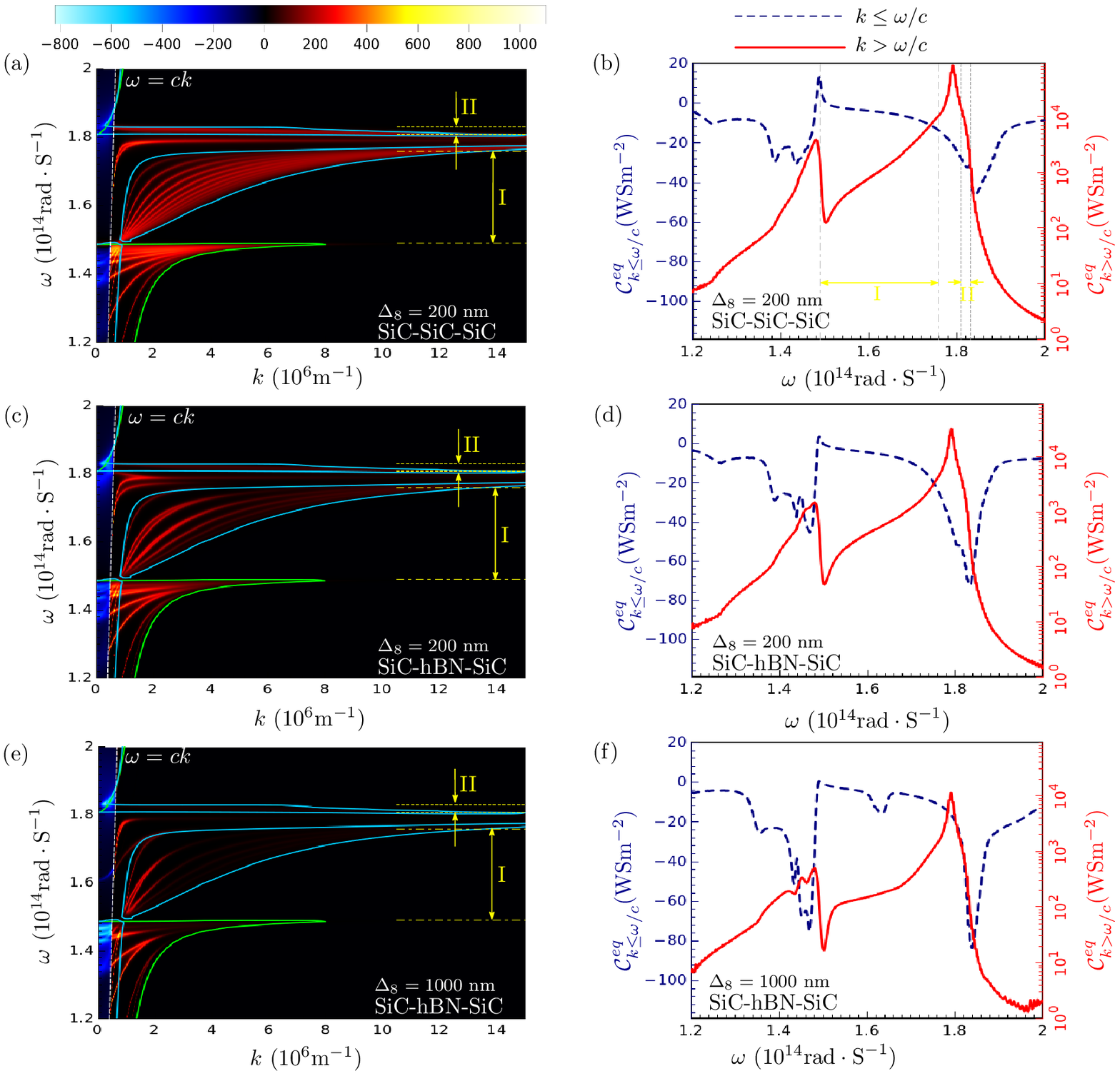} 
\caption{(Color online) Left panel: The steady-state thermal current dispersion relation $\mathcal{S}^{eq}$ in $(\omega,k)$ space fixing $\delta=100$nm and $\Delta_{i\neq 8}=200$nm for boundary condition ($T_L,T_1,T_{15},T_R$)=($300,400,300,300$)~K. The yellow lines mark the hyperbolic bands as determined from effective medium theory. (a) SiC barrier, $\Delta_8=200$nm. (c) hBN barrier, $\Delta_8=200$nm.  (e) hBN barrier, $\Delta_8=1000$nm. 
The solid blue (green) curves  are the borders of the Bloch bands for  P-polarized (S-polarized) modes, respectively. Right panel: The monochromatic thermal current coefficient $\mathcal{C}^{eq}(\omega)$ of propagating ($k<\omega/c$) and evanescent ($k>\omega/c$) modes for the same structure in the left panel:(d) SiC barrier, $\Delta_8=200$nm. (e) hBN barrier, $\Delta_8=200$nm.  (f) hBN barrier, $\Delta_8=1000$nm.  
} 
 \label{Figure.5}
\end{figure*}
In order to see the structure of contributing modes, we have investigated the steady-state thermal current dispersion relation $\mathcal{S}^{eq}$ in $(\omega,k)$ space for both the SiC-SiC-SiC and SiC-hBN-SiC structures in Fig.~\ref{Figure.5}. We start with the structure investigated Fig.~\ref{Figure.2}(a) which is made of alternating layers of SiC and vacuum. The geometrical parameters in Fig.~\ref{Figure.5}(a) are $\Delta_8=\Delta=200$nm for the SiC layers and $\delta=100$nm for the vacuum gaps. Such a periodic structure behaves as a photonic crystal and can be described by homogeneous anisotropic media
with the effective permittivities $\epsilon_{\perp}=f_1\epsilon_1+f_2\epsilon_2$ and $\epsilon_{ \parallel}=(\epsilon_1\epsilon_2)/(f_1\epsilon_2+f_2\epsilon_1)$, where $\epsilon_1$ and $\epsilon_2$ are the permitivities of the vacuum gap and SiC layer respectively. Based on the choice of parameters in Fig.~\ref{Figure.5}(a), the filling fraction are $f_1=\frac{\delta}{\delta+\Delta}=\frac{1}{3}$ and $f_2=\frac{\Delta}{\delta+\Delta}=\frac{2}{3}$, respectively. The effective permittivities are helpful for determining the hyperbolic frequency bands of the structure fulfilling the condition $\epsilon_{\perp}\epsilon_{ \parallel}<0$. The  dashed yellow lines in Fig.~\ref{Figure.5} represent the edges of the type I $(\epsilon_{\perp}>0,~\epsilon_{ \parallel}<0)$ and type II $(\epsilon_{\perp}<0,~\epsilon_{ \parallel}>0)$ hyperbolic modes determined from effective medium theory. Moreover, these structures can suppurt the bloch modes which are located in Bloch bands. The borders of the Bloch bands for S- and P-polarized waves can be easily determined for infinite multilayer structure and semi-infinite materials \cite{doi:10.1063/1.3385156,doi:10.1063/1.4800233,PRYAMIKOV20111314}. For the $|\textrm{SiC}|\textrm{vacuum}|$ periodic structure the Bloch modes are fulfilling the dispersion relation 
\begin{equation}
\begin{aligned}
\cos\left(\beta_{B}(\delta+\Delta)\right)= & -\xi\sin(\beta_1\delta)\sin(\beta_2\Delta) \\
 &+\cos(\beta_1\delta)\cos(\beta_2\Delta),
\label{eq16}
\end{aligned}
\end{equation}
with
\begin{eqnarray}
\label{eq17}\xi&=&\frac{1}{2}\left(\frac{\epsilon_1\beta_2}{\epsilon_2\beta_1}+\frac{\epsilon_2 \beta_1}{\epsilon_1\beta_2}\right)~~\textrm{for P-polarized modes},\\
\label{eq18}\xi&=&\frac{1}{2}\left(\frac{\beta_1}{\beta_2}+\frac{\beta_2}{\beta_1}\right)~~~~~~~~\textrm{for S-polarized modes}.
\end{eqnarray}
Here, $\beta_{i=1,2}=\sqrt{\epsilon_i\omega^2/c^2-k^2}$ are the wave-vectors  along the optical axis of the structure in each layer and $\beta_B$ is the Bloch wave-vector inside the periodic structure.

The solid blue (green) curves in Fig.~\ref{Figure.5}(a), (c) and (d) are the borders of the Bloch bands for  P-polarized (S-polarized) modes, respectively. From the dispersion relation of the SiC-SiC-SiC shown in Fig.~\ref{Figure.5}(a), it is apparent that the thermal current has a quite large contributions stemming from modes inside  Bloch bands. However, there is also a contribution of narrow-band resonance modes outside the Bloch regions and concentrated around the surface phonon polariton resonance of the SiC layer at $\omega_{\text{SPhP}}=1.787\times 10^{14}$ rad/s. It is interesting to know that according to the values selected for the thermal boundary conditions ($T_L,T_1,T_{15},T_R$)=($300,400,300,300$)~K, the direction of thermal current  $J^{eq}$ is from left to right (i.e. positive). However, the sign of $\mathcal{S}^{eq}(\omega,k)$ is not necessarily positive. On the other hand, some modes have positive contributions to the thermal current while the rest have negative contributions. Since the slabs are very close to each other $\delta=100$nm, we observe in Fig.~\ref{Figure.5}(a) that the contribution of evanescent  waves in thermal current is positive in $(\omega,k)$ space, while the contribution of propagating waves is mostly negative. The contribution of propagating and evanescent  modes is plotted in Fig.~\ref{Figure.5}(b) where we show the monochromatic thermal current coefficient $\mathcal{C}^{eq}(\omega)$. From this figure, it becomes obvious that the thermal current is dominated by positive contribution of evanescent modes in  both the Bloch and Non-Bloch regions. However, the negative contribution of propagating waves are dominated solely by modes located inside the Bloch bands.

If we introduce a defect for the chosen structure,  by changing the width of the middle layer or by changing the composition of its material, the structure will no longer be periodic. Such a defect can introduce multiple defect states in the band gap that have different frequencies \cite{doi:10.1063/1.1898450}. We start by the case where the SiC barrier is replaced with hBN material with same thicknesses. The temperature profile of this structure is investigated in Fig.~\ref{Figure.2}(b). The thermal current dispersion relation of this structure is shown in Fig.~\ref{Figure.5}(c). From this figure, one notices that the surface Bloch modes in frequencies inside the hyperbolic band type II is fully suppressed in this case. Moreover, we observe that a significant decrease in both the intensity and number of confined Bloch modes in polarization TE (green region) and TM (blue region) is accompanied by the emergence of localized P-polarized modes in the photonic band gap of the structure.

From the shown plot in Fig.~\ref{Figure.5}(e), it is  apparent that the remaining evanescent modes in the Bloch bands of SiC-SiC-SiC structure, persist when we increase the hBN barrier width to $\Delta_8=1000$nm. However, their contribution slightly suppressed. We also notice that the intensity of the two defect modes drastically decreased, and they are shifted  away from the $\omega_{\textrm{\tiny{SPhP}}}$ of SiC.
In contrary, we observe the amplification in the  propagating modes with negative contribution to thermal current.  Moreover, new channels for the opposing thermal currents are also emerged in the spectral frequency window between $\omega_T$ and $\omega_L$ of the SiC layer. Also, in Fig.~\ref{Figure.5}(f), we see that the contribution of evanescent waves to the monochromatic thermal current is negative in all frequency ranges. As a result, in addition to the significant change in the dispersion curves of photons, the increase in the opposing thermal currents in $\mathcal{S}^{eq}(\omega,k)$ is responsible for the increase in the thermal resistance  for large barrier thicknesses.

\begin{figure}
\includegraphics[height=6cm,width=8cm]{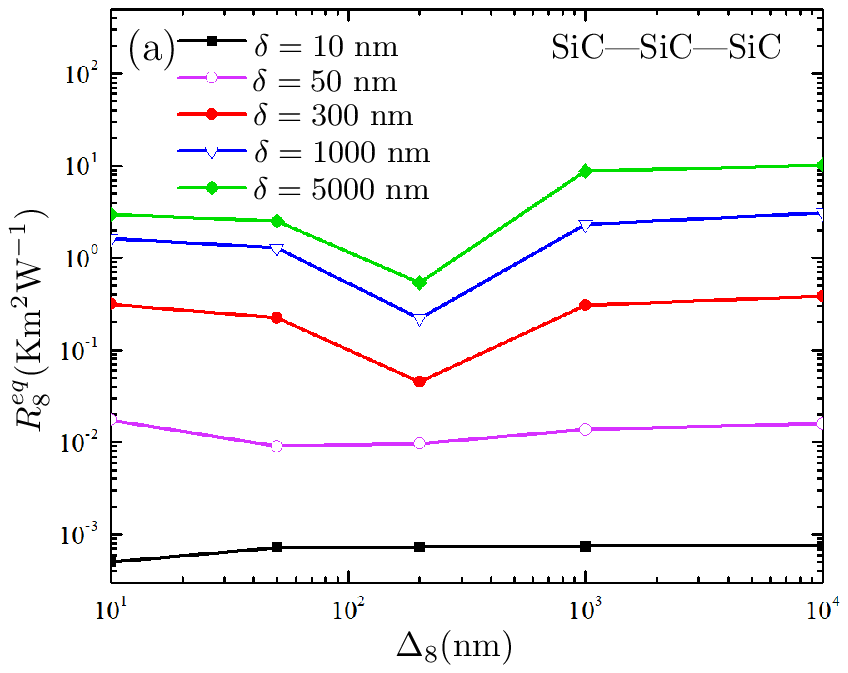}
\includegraphics[height=6cm,width=8cm]{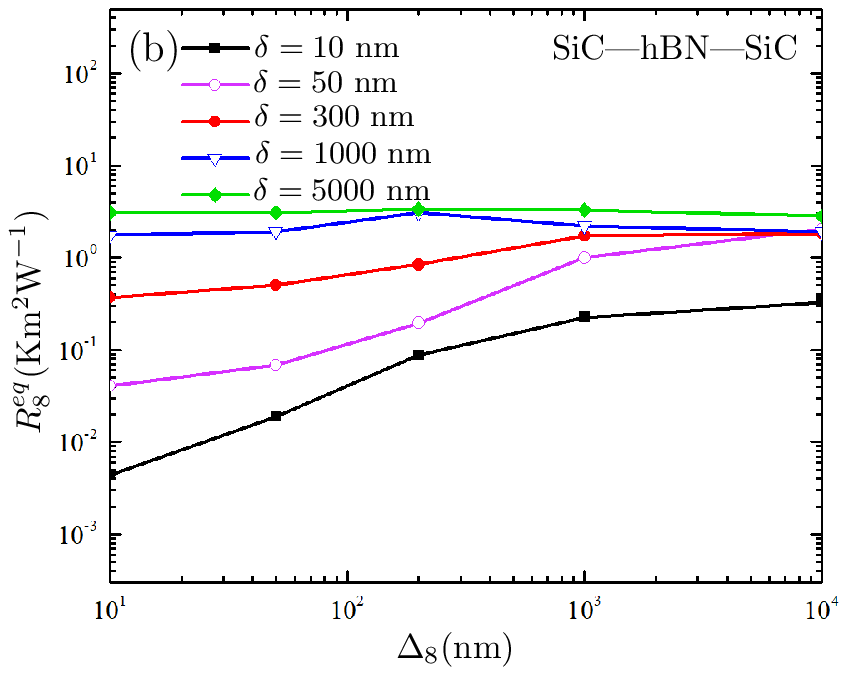}
\caption{(Color online) The steady state linear resistance of the thermal barrier as a function of barrier thickness for different vacuum gap widths values. The thickness of other slabs are the same and considered $\Delta=200$nm. The material used for the barrier is: (a) SiC, (b) hBN.} 
 \label{Figure.6}
\end{figure}

To compare the thermal resistance of hBN with SiC barrier, we have calculated the linear resistance of barriers with different thicknesses. Shown  in  Fig.~\ref{Figure.6}(a) are our results for the resistance of SiC barrier as a function of the barrier thickness, for different vacuum gap separation distances. The thickness of the other slabs in the system are $\Delta=200$nm, as in the previous figures. It is striking that, for a given width for vacuum gaps, the calculated resistance depends not only on the thickness of the barrier but on the thickness of the other slabs too. It is interesting that this dependence follows a certain rule for SiC-SiC-SiC structure. For large separation distances $\delta \geqslant \Delta$, the thermal resistance of the barrier of width $\Delta_8 \simeq \Delta$ is always minimal. 

To understand the physics behind this result, we note that as discussed earlier, for very small separation distances between SiC layers, the thermal current is dominated solely by the evanescent modes of the SiC layers. For sufficiently large separation distances, the thermal current is mainly due by the contributions of propagating modes. In such a case, the distance between the leftmost  layer (left reservoir) and the rightmost (right reservoir) layer  are far enough and since $T_1>T_N$ and $T_N=T_R=T_L$, we may approximate the thermal current passing through the system by radiation energy emitted from the left reservoir that transmitted through the periodic structure made by alternating layers of a SiC and a vacuum gap, say $|n_1|n_2|n_1|n_2|\cdots|n_1|n_2|$ bounded on both sides by vacuum, where $n_1$ and $n_2$ are the refractive index of the SiC layers and vacuum gaps respectively. Such a periodic structure can increase the transfer of the incident thermal current impinging from the left, by decreasing the reflected thermal current, similar to anti-reflection structures\cite{jac}. The condition of excitation of the {\it anti-reflecting modes}  satisfies in frequency ranges where the optical thickness of each layer equals to a quarter of the incident wavelength of the thermal current. In general the number of layers and the intrinsic properties of each
layer can be engineered so that the excitation of anti-reflecting modes, which produce destructive interference in the beams reflected from the interfaces and constructive interference in the corresponding transmitted beams, occurs within the desired frequency range. For periodic structure used here, the contribution of these mode to the heat transfer becomes important since they are excited by the incident radiation from the left reservoir, located around the wien frequency at $T_1 \geqslant T_B=300~$K. However, by perturbing the periodicity (i.e. $\Delta_8\neq\Delta$), the anti-reflecting modes are shifted away from the Wien’s frequency of the left reservoir so that they do not contribute anymore to radiative heat transfer. As a result, it reduces the thermal current, which is equivalent to increase in the thermal resistance.

On the other side, as the separation distance decreases, this minimum occurs for smaller barrier thicknesses, i.e. $\Delta_8\ll\Delta$.  The conditions of the hBN barrier is quite different from those of SiC, see Fig.~\ref{Figure.6}(b). In this case, increasing the barrier thickness is accompanied by an increase in thermal resistance. Although this increase is negligible and slightly oscillatory for high separation distances, but similar to the SiC barrier, the resistance is saturated at large barrier thicknesses, as expected. 

\begin{figure}
\includegraphics[height=6cm,width=8cm]{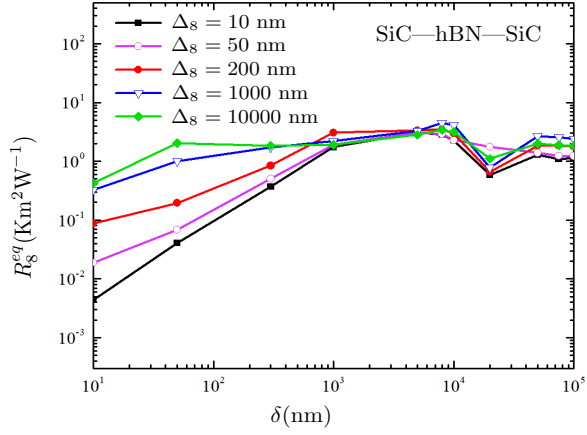}
\caption{(Color online) The variation of the barrier resistance as a function of slab separation distance in the SiC--hBN--SiC system. The results are shown for different barrier thickness values and the thickness of SiC slabs are the same and considered $\Delta=200$nm.}
\label{Figure.7}
\end{figure}
\subsection{VACUUM GAP EFFECT}\label{sec3d}
Finally, the influence of the width of the vacuum gaps on the linear resistance of the hBN barrier is presented in Fig.~\ref{Figure.7}. The steady-state linear resistance is shown for different barrier thickness values, as a function of vacuum gap widths from near-field to far-field regime. It is seen that the thermal resistance of the hBN barrier and consequently the thermal resistance of the whole structure in the near field regime is much lower than the far field regime. On the other hand, the rapid decrease in the transmission probabilities between slabs by increasing the vacuum gap widths is responsible for the power law increment of the thermal resistance in the near-field regime which is modulated by logarithmic periodic oscillations and saturates in far-field limit. In agreement with previous results, the barrier resistance increased for larger barrier thicknesses in near-field regime.

\section{Conclusion}\label{sec4}
In summary, we have used an electrical circuit approach to introduce heat resistance for the components of systems that transmit energy through radiation at the nanoscale. For this purpose we proposed a method for calculating the steady-state radiative current in parallel planar systems. This method can be a useful guide for understanding and optimizing the thermal performance of nanoscale systems. The simulation results indicate that the temperature profile in a parallel planar system, exhibits fantastic changing characteristic around the barriers. We have shown the thermal insulation occurs due to the presence of thermal barrier and the temperature does not show a monotonic trend across the barrier, instead, there is a temperature difference athwart the barrier.  
 \appendix
\numberwithin{equation}{section}
\makeatletter
\renewcommand{\theequation}{\Alph{section}.\arabic{equation}} 
 \section{Transmission coefficients in parallel planar systems}\label{sec5}
 The system under consideration consist of $N$ parallel slabs placed at $z_i$ along the z-axis. The separation distance between the consecutive slabs $i$ and $i+1$ is $\delta_i=z_{i+1}-z_i-\Delta_{i}/2-\Delta_{i+1}/2$ where  $\Delta_i$ is thickness of the $i$-th slab. The first ($i=1$) and the last ($i=N$) slabs  are connected to reservoirs with fixed temperatures $T_1$ and $T_N$, respectively. The indexes $i=0\equiv L$ and $i=N+1\equiv R$ are used for the left and the right thermal baths which are kept at fixed temperatures $T_L$ and $T_R$, respectively. The many-body energy transmission coefficients $\mathcal{T}^{j,i}$ take into account the presence of different slabs at the same time and can be fully determined in terms of $\hat{\mathcal{T}}=\hat{\mathcal{T}}(\omega, k ,p)$\cite{4bodydynamics}:

\begin{equation}\begin{split}
&\mathcal{T}^{L, i} = \hat{\mathcal{T}}^{L}_{ i -1} - \hat{\mathcal{T}}^{L}_{ i },\\
&\mathcal{T}^{ j , i } = \hat{\mathcal{T}}^{ j }_{ i -1} -\hat{\mathcal{T}}^{ j -1}_{ i -1} - \hat{\mathcal{T}}^{j }_{ i } +\hat{\mathcal{T}}^{ j -1}_{i },\\
&\mathcal{T}^{R, i } = -\hat{\mathcal{T}}^{N}_{ i -1} + \hat{\mathcal{T}}^{N}_{ i },
\end{split}\label{rel_trans_coeff}\end{equation}
for $j,i=1,\cdots,N$. The definition of these coefficients are as follow:

\begin{widetext}
\begin{equation}\begin{split}
\hat{\mathcal{T}}^{ j }_{\gamma}
&= 
\frac{\Pi^{\text{pw}} \big|\tau^{ j +1\to\gamma}\big|^2 \big(1 -\big|\rho_+^{L\to j }\big|^2\big)
\big( 1 -\big|\rho_-^{\gamma+1\to R}\big|^2 \big)}
{\big|1- \rho_+^{L\to\gamma} \rho_-^{\gamma+1\to R} \big|^2 
\big|1- \rho_+^{L\to j } \rho_-^{ j +1\to \gamma}\big|^2} 
+ 
\frac{\Pi^{\text{ew}} 4\big|\tau^{ j +1\to\gamma}\big|^2\text{Im}\big(\rho_+^{L\to j }\big)
\text{Im}\big(\rho_-^{\gamma+1\to R}\big)}
{\big|1- \rho_+^{L\to\gamma} \rho_-^{\gamma+1\to R} \big|^2 
\big|1- \rho_+^{L\to j } \rho_-^{ j +1\to \gamma}\big|^2},
\qquad  j <\gamma,\\
\hat{\mathcal{T}}^{\gamma}_{\gamma}
&=    
\frac{\Pi^{\text{pw}} \big(1 -\big|\rho_+^{L\to\gamma}\big|^2\big)
\big( 1 -\big|\rho_-^{\gamma+1\to R}\big|^2 \big)}
{\big|1- \rho_+^{L\to\gamma} \rho_-^{\gamma+1\to R}\big|^2}
+ 
\frac{\Pi^{\text{ew}} 4\text{Im}\big(\rho_+^{L\to\gamma}\big)
\text{Im}\big(\rho_-^{\gamma+1\to R}\big)}
{\big|1- \rho_+^{L\to\gamma} \rho_-^{\gamma+1\to R}\big|^2},\\
\hat{\mathcal{T}}^{ j }_{\gamma}
&=
\frac{\Pi^{\text{pw}} \big|\tau^{\gamma+1\to j } \big|^2 
\big(1-\big|\rho_+^{L\to\gamma}\big|^2\big)
\big(1-\big| \rho_-^{ j +1\to R}\big|^2 \big)}
{\big|1- \rho_+^{L\to j } \rho_-^{ j +1\to R}\big|^2
\big|1- \rho_+^{L\to\gamma} \rho_-^{\gamma+1\to  j }\big|^2}
+
\frac{\Pi^{\text{ew}} 4\big|\tau^{\gamma+1\to j } \big|^2 
\text{Im}\big(\rho_+^{L\to\gamma}\big)
\text{Im}\big(\rho_-^{ j +1\to R}\big)}
{\big|1- \rho_+^{L\to j } \rho_-^{ j +1\to N+1}\big|^2
\big|1- \rho_+^{L\to\gamma} \rho_-^{\gamma+1\to  j }\big|^2}.
\qquad  j >\gamma,
\end{split}\label{coeff_seq}\end{equation}
\end{widetext}
These coefficients satisfy the reciprocity relation $\hat{\mathcal{T}}^{ j }_{i}=\hat{\mathcal{T}}^{i}_{ j }$. The many-body scattering coefficients $\rho_+^{ j \to m}$, $\rho_-^{ j \to m}$ and $\tau^{j\to m}$ are given by
\begin{equation}\begin{split}
\rho_+^{ j \to m}&= \hat{\rho}_+^{ j \to m}e^{-i k_z\left(\Delta_m+2z_m\right)},\\
\rho_-^{ j \to m}&= \hat{\rho}_-^{ j \to m}e^{-i k_z\left(\Delta_j-2z_j\right)},\\
\tau^{ j \to m}&= \hat{\tau}^{ j \to m}\exp\Bigl(-i k_z \sum_{\ell=j}^m \Delta_\ell\Bigr),
\end{split}\label{hat_coefficients}\end{equation}
where
\begin{equation}\begin{split}
\hat{\rho}_+^{ j \to m}&= \rho_m + (\tau_m)^2 \hat{\rho}_+^{ j \to m-1}u^{ j \to m-1, m} e^{2i k_z \delta_{ m-1}} ,\\
\hat{\rho}_-^{ j \to m}&= \rho_j + (\tau_j)^2 \hat{\rho}_-^{ j +1\to m}  u^{ j , j +1\to m} e^{2i k_z\delta_j }, \\
\hat{\tau}^{ j \to m}&= \hat{\tau}^{ j \to m-1} u^{ j \to m-1, m}  \tau_m,
\end{split}\end{equation}
with 
\begin{equation}\begin{split}
u^{ j \to m-1, m}&= \left(1-\hat{\rho}_+^{ j \to m-1} \rho_m e^{2i k_z\delta_{ m-1}} \right)^{-1},\\
u^{ j , j +1\to m}&= \left(1-\rho_j \hat{\rho}_-^{ j +1\to m} e^{2i k_z\delta_j}\right)^{-1}.
\end{split}\end{equation}
Here, $\rho_ j $ and $\tau_ j $ are the scattering  coefficient for a single body which are given by

\begin{equation}\begin{split}
\rho_{ j }&=r_{p, j }\frac{1-e^{2i k_{z  j }\Delta_j }}{1-r^2_{p, j }
e^{2i k_{z  j }\Delta_j }},\\
\tau_{ j }&=\frac{\left(1- r_{p, j }^2\right)
e^{i k_{z  j }\Delta_j }}
{1-r^2_{p, j }e^{2i k_{z  j }\Delta_j}}.
\end{split}\end{equation}
In equation (A6), $r_{p,j}$ is the Fresnel coefficient in which $p$ indicates the polarization. This coefficient for two possible polarizations including TE and TM, is defined as
\begin{equation}
r_{\mathrm{TE},  j }=\frac{\mu_ j  k_z-k_{z  j }}{\mu_ j  k_z+k_{z  j }}, \qquad
r_{\mathrm{TM},  j }=\frac{\epsilon_ j  k_z-k_{z  j }}{\epsilon_ j  k_z+k_{z  j }}.
\label{Fresnel_coefficients}
\end{equation}
In the above equations $\epsilon_ j $ and $\mu_ j $ are electric permittivity and magnetic permeability, representing the optical properties of $ j $th slab.
 
\section*{reference}
%
\end{document}